# Density-Accelerated Relaxation in Active–Passive Colloidal Networks with Competing Interactions

Pronay Dutta,[1,†] and Hima Nagamanasa Kandula[1,*]

[1]Department of Physics, University of Massachusetts Amherst, Amherst, MA 01003, USA

We investigate quasi-2D active-passive colloidal mixtures with electric field-mediated non-identical interactions: unlike species attract and like species repel. Mapping the steady states across total area fraction and active particle area fraction identifies the regime of active percolated network formation among active fluid and clustered states. Within the percolated network regime, relaxation speeds up with increasing density, contrary to expectations for dense particulate gels. Particle-resolved dynamics and microstructure link this acceleration to activity-driven rearrangements enabled by asymmetric interactions.

Colloidal gels are space-spanning percolated networks [1] formed by the aggregation of particulate matter [2]. They are invaluable due to their wide use in industries from food to pharmaceuticals, where their mechanical properties are critical to their performance [3,4]. Colloidal gels also serve as an ideal model system for gaining microscopic insights into the observed macroscopic mechanical properties [5–7]. However, once formed, particulate gels typically become dynamically arrested, limiting the tunability of their structure and hence mechanical response in situ. Active colloids offer a promising route to overcome this limitation by continuously injecting energy at the particle scale. However, most active colloidal systems display dynamic phase separation rather than achieving condensed states like space-spanning networks [8–11]. A possible strategy is to combine active and passive particles. Existing experiments have introduced a small fraction of active dopants into a passive gel [12,13], but these gels are assembled in a passive state and then activated. This differs from biological networks, where networks form in the presence of activity, allowing activity to participate in network selection. Simulations suggest that introducing attractive interactions in active–passive mixtures can generate diverse steady states, including networks [14–16]. Yet, experimental realizations of active-passive mixtures that form percolated networks with tunable dynamics remain scarce. Specifically, how particle interaction and activity jointly govern relaxation in percolated networks remains an outstanding challenge.

Electric field-driven dipolar interactions provide a means to tune inter-particle interactions reversibly [17–19]. Field-driven active colloids allow propulsion and interparticle interactions to be controlled through a common external handle [20,21]. E-field has been shown to produce chains in passive dielectric colloids and diverse organizations in 3D systems of active colloids [22–25]. However, limited single-particle resolution in existing 3D studies constrains mechanistic understanding, while their 2D counterparts lack in-plane attractive interactions needed to form rich network states. Incorporating passive components into 2D active systems can introduce effective interparticle attractions, while preserving particle-level access. Such a platform enables direct quantification of the role of interaction asymmetry and activity in controlling the steady-state structure and relaxation dynamics.

In this letter, we report steady states formed by mixtures of active Janus and passive silica colloids subjected to an orthogonal alternating electric field (AC E-field), in the phase space of total area fraction ($\emptyset_T$) and the active particle area fraction ($\frac{\emptyset_A}{\emptyset_T}$) [26]. The interspecies dipolar attraction seeds clustering and network growth, while the like-species repulsion prevents collapse into dense aggregates and stabilizes percolated networks. Surprisingly, within the network regime, relaxation accelerates with increasing particle density. We show that this behavior arises from activity–interaction coupling, which drives local rearrangements within the network. These results establish a platform for tuning structure and relaxation time in particulate networks.

A binary suspension of active and passive colloidal particles is sandwiched between indium tin oxide-coated (ITO) coverslips and an alternating electric field, a square wave of $E_0 = 0.06V = 1\,MHz$, is applied (**Fig 1a**). 3 µm diameter silica spheres are used for passive colloids. To make active colloids, we deposit 30 nm of titanium on one hemisphere of the passive silica colloids to make them Janus, which, when subjected to an alternating electric field, self-propel [27](**SI 1**). At $f = 1\,MHz$, in the 2D plane of observation, active and passive particles attract each other, while particles of the same species repel (**Fig 1b**, **SI Mov. 2**).

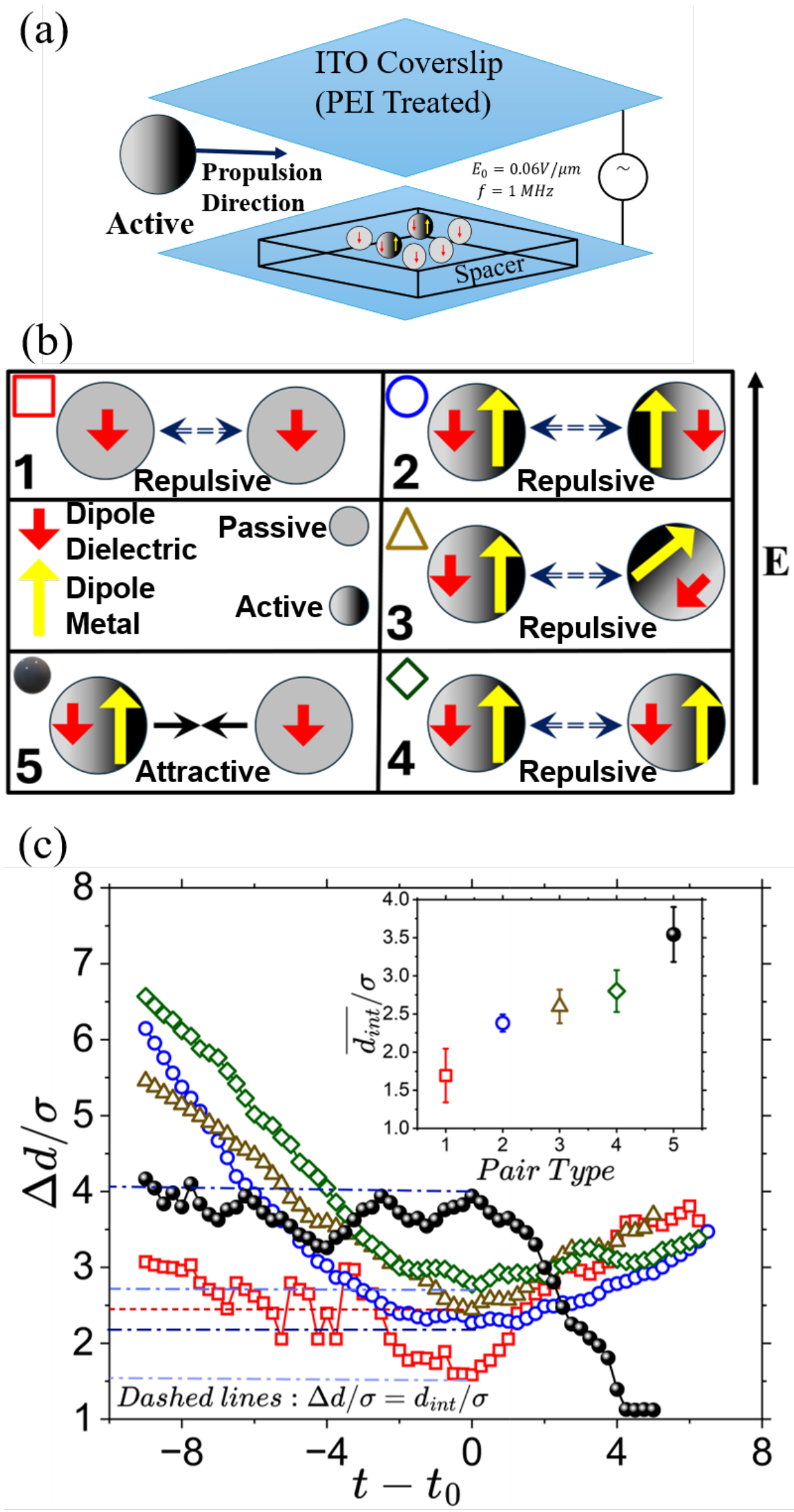

**Figure 1:** Pair interaction between unlike species(active-passive) and like species (active-active and passive-passive). **(a)** Schematic of the experimental setup**. (b)** List of possible pairwise interactions. The arrows correspond to the instantaneous dipole moment. The length of the arrows represents dipole strength. The symbols and the numbers correspond to the data in c for the corresponding particle pairs. **(c)** $\Delta d\ vs\ (t - t_0)$ for a single pair for each type of interaction. Dashed lines represent the range of significant interaction($d_{int}$) of a specific pair. **Inset:** Mean range of significant interaction ($\overline{d_{int}}$) for particle pairs. $\overline{d_{int}}$ is obtained by averaging ~30 pairs of each type. All lengths in c are normalized by particle diameter $\sigma$.

The interactions between particles arise from field-induced dipoles and electrostatic effects. Active and passive colloid attraction originates from oppositely oriented induced dipoles on the metallic hemisphere of the active colloid and the dielectric passive particle, respectively. Passive–passive interactions are repulsive due to like-induced dipoles. Active particle pairs have repulsive interactions for all approach orientations, despite the Janus character. We attribute this to the Polyethyleneimine (PEI) coating, which imparts a large positive zeta potential (~ +40 mV) and suppresses dipolar attraction between metallic and dielectric hemispheres [28] [29] (**SI 2,SI Mov. 1**). Without PEI, active particles typically assemble into chains at 1 MHz, as widely reported. The surface modification of the colloids with PEI enables asymmetric interactions between non-identical pairs.

We quantify the inter-particle interactions by tracking isolated particle pairs at low total area fraction ($\phi_T \sim 0.01$). We extract the separation distance between particle pairs, $\Delta d(t)$, with $t_0$ marking the time for the onset of interaction. Beyond $t_0$, unlike pairs (active–passive), show a rapid decrease in $\Delta d$, whereas like pairs (active–active and passive–passive) show increasing $\Delta d$ (**Fig 1c; SI 2; SI Mov. 2**) [30,31]. We define the mean interaction distance $\overline{d_{int}}$ as the separation at which particle motion deviates from free diffusion. Therefore $\overline{d_{int}}$ corresponds to the distance at which interaction forces dominate thermal fluctuations. Consequently, larger $\overline{d_{int}}$ reflects stronger interactions. We find that $\overline{d_{int}}$ is largest for active–passive attraction, smaller for active–active repulsion, and smallest for passive–passive repulsion (**Fig 1c inset**). The observed hierarchy in interaction strength arises from the interplay between activity and induced dipole moments. Active particles, owing to their metallic hemisphere, develop larger dipole moments than silica particles. However, active-active interactions are not the strongest due to self-propulsion, which transiently drives particles into closer proximity, partially offsetting repulsive interactions and reducing the measured $\overline{d_{int}}$. These results provide a direct measure of the complex interparticle interactions present in the system.

The hierarchy in interaction strength, combined with the asymmetric nature of the interactions, promotes preferential binding of an active particle with multiple passive particles (typically 3–4) (**SI Fig 2a**), while suppressing active–active nearest-neighbor contacts. These locally preferred active–passive configurations define the structural basis of the steady states observed across active and passive area fractions (**Fig 2a-2c).** Importantly, this mechanism distinguishes the states reported here from the dynamic clustering or motility-induced phase separation commonly observed in active–passive mixtures.

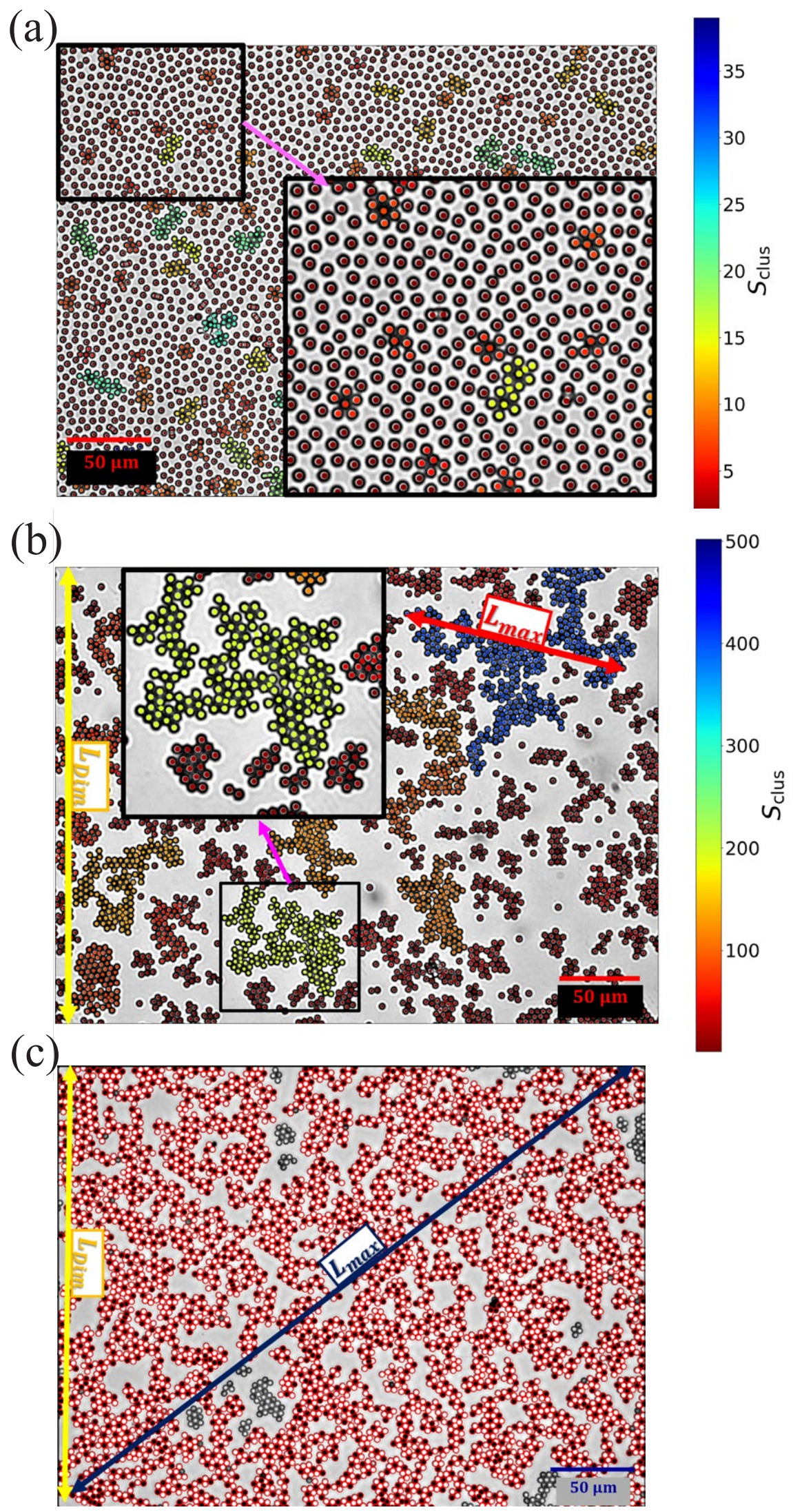

**Figure 2:** Observed steady-states. **(a)** fluid state ($\emptyset_T \approx 0.42; \frac{\emptyset_A}{\emptyset_T} \approx 0.1$). **(b)** cluster state ($\emptyset_T \approx 0.42; \frac{\emptyset_A}{\emptyset_T} \approx 0.30$) In a and b, particles are color-coded by the cluster size, $S_{Clus}$. **(c)** percolated network state ($\emptyset_T \approx 0.56; \frac{\emptyset_A}{\emptyset_T} \approx 0.35$). Particles in the biggest cluster are highlighted in red. The particles colored black and white represent active and passive ones, respectively.

We classify the observed states using the cluster size distribution, $P(S_{clus})$. Here, $S_{clus}$, is the number of particles in a cluster. A state is identified as fluid when more than 50% of the particles exist as monomers, and as a cluster state if more than 50% exist as dimers and beyond (**Fig 3a**). With increasing total area fraction, $\emptyset_T$, the largest cluster grows to span the entire field of view, indicating the emergence of a percolated network state. We define this state by requiring that the maximum separation between two particles in the largest cluster ($L_{max}$) exceeds the size of the field-of-view($L_{Dim}$) (**Fig 2c,3b;SI 3**) [32]. Equivalently, a cluster state is classified as a network state if $\frac{L_{max}}{L_{Dim}} \geq 1$.

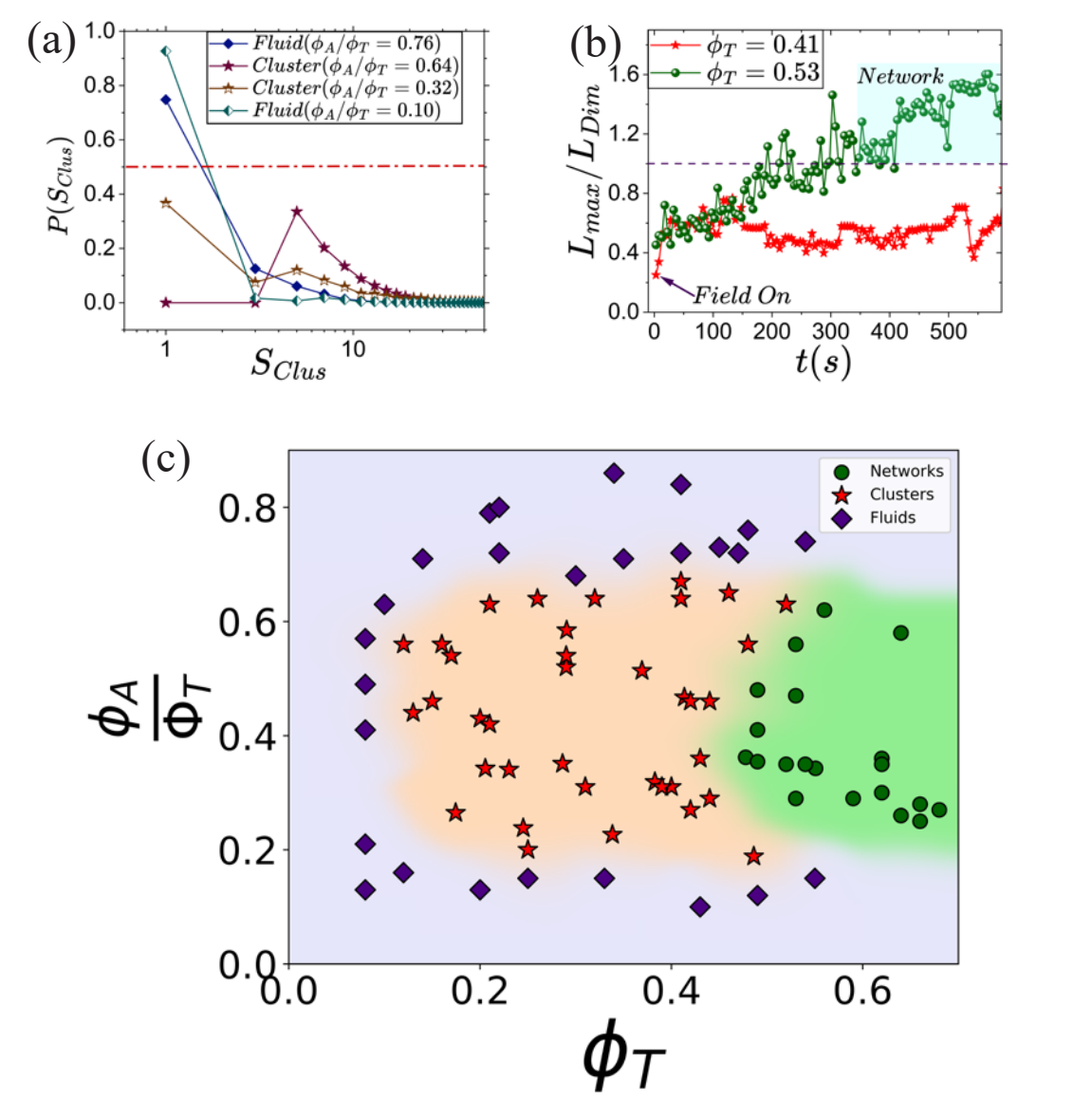

**Figure 3:** State diagram. **(a)** $P(S_{Clus})$ for fluid and cluster states: $\emptyset_T \approx 0.42$ and varying $\frac{\emptyset_A}{\emptyset_T}$ . The red dashed line indicates a probability of 50%. **(b)** $\frac{L_{max}}{L_{Dim}}$ vs time of the largest cluster in the cluster (red star) and percolated network states (green sphere) for $\frac{\emptyset_A}{\emptyset_T} \approx 0.47$ **(c)** Steady states $\emptyset_A/\emptyset_T$ - $\emptyset_T$ phase space. The colors in the background are guides to the eye, with fluids in purple, cluster states in red, and networks in green. The symbols correspond to the experimental data.

**Fig 3c** maps the state diagram, identifying fluid, cluster, and network regimes in $\frac{\emptyset_A}{\emptyset_T}$ - $\emptyset_T$ phase space. Here $\emptyset_T = \emptyset_A + \emptyset_P$, $\emptyset_A$ and $\emptyset_P$ correspond to the active and passive particle area fractions, respectively. The state diagram shows a clear progression from fluids to clusters to percolated networks with increasing $\emptyset_T$ over a range of $\frac{\emptyset_A}{\emptyset_T}$. In contrast, increasing $\frac{\emptyset_A}{\emptyset_T}$ for a given $\emptyset_T$, yields a non-monotonic transition from fluid to clustered/network state and back to fluids. This non-monotonic behavior arises from the competing interactions: at lower $\frac{\emptyset_A}{\emptyset_T}$, active-passive attraction promotes clustering, while at higher $\frac{\emptyset_A}{\emptyset_T}$, the reduced number of passive particles limits binding, and active-active repulsion becomes dominant, restoring a fluid state (**SI 4, SI Mov. 3**). Ensemble-averaged MSD shows that activity persists across fluid, cluster, and network states, with super-diffusive to nearly ballistic motion in $\frac{\emptyset_A}{\emptyset_T}$ - $\emptyset_T$phase space. **(EM Fig 1)** [33].

Having identified the phase space of networks, we compare their dynamics via the characteristic relaxation time, $\tau$, extracted using the ensemble-averaged self-intermediate scattering function F(k,Δt) [34]. The decay of $F(k, \Delta t)$ is fit to an exponential ($exp\left[-\left(\frac{t}{\tau}\right)\right]$) or stretched exponential ($exp\left[-\left(\frac{t}{\tau}\right)^{\beta}\right]$) to extract $\tau$ [35,36]. **Fig 4a** shows $F(k, \Delta t)$ for three $\frac{\emptyset_A}{\emptyset_T}$at a fixed $\emptyset_T$, demonstrating faster decay with increasing $\frac{\emptyset_A}{\emptyset_T}$. **Fig 4b** shows $F(k, \Delta t)$ for varying $\emptyset_T$ at fixed $\frac{\emptyset_A}{\emptyset_T}$, demonstrating a faster decay with increasing $\emptyset_T$, a counterintuitive behavior. The extracted relaxation times for all the network states show that $\tau$ decreases with increasing $\emptyset_T$ for various $\frac{\emptyset_A}{\emptyset_T}$, establishing a robust increase in dynamics at higher particle number density (**Fig 4c, SI Mov. 4**).

To understand the origin of the faster relaxations in networks at higher $\emptyset_T$, we examine their microstructure. We focus on two representative network states at $\frac{\emptyset_A}{\emptyset_T} \approx 0.35$ : a slow network at $\emptyset_T \approx 0.49$ with $\tau > 500s$, and a fast network at $\emptyset_T \approx 0.62$, with $\tau \sim 3s$ (states highlighted by circles in Fig 4c).

We first analyze the probability distribution of the number of nearest neighbors $P(N_{nn})$ within the system spanning cluster. Here, $N_{nn}$ is the number of nearest neighbors per particle. The peak of $P(N_{nn})$ shifts to larger values with increasing $\emptyset_T$, indicating that the networks become more compact (**Fig 5a**).

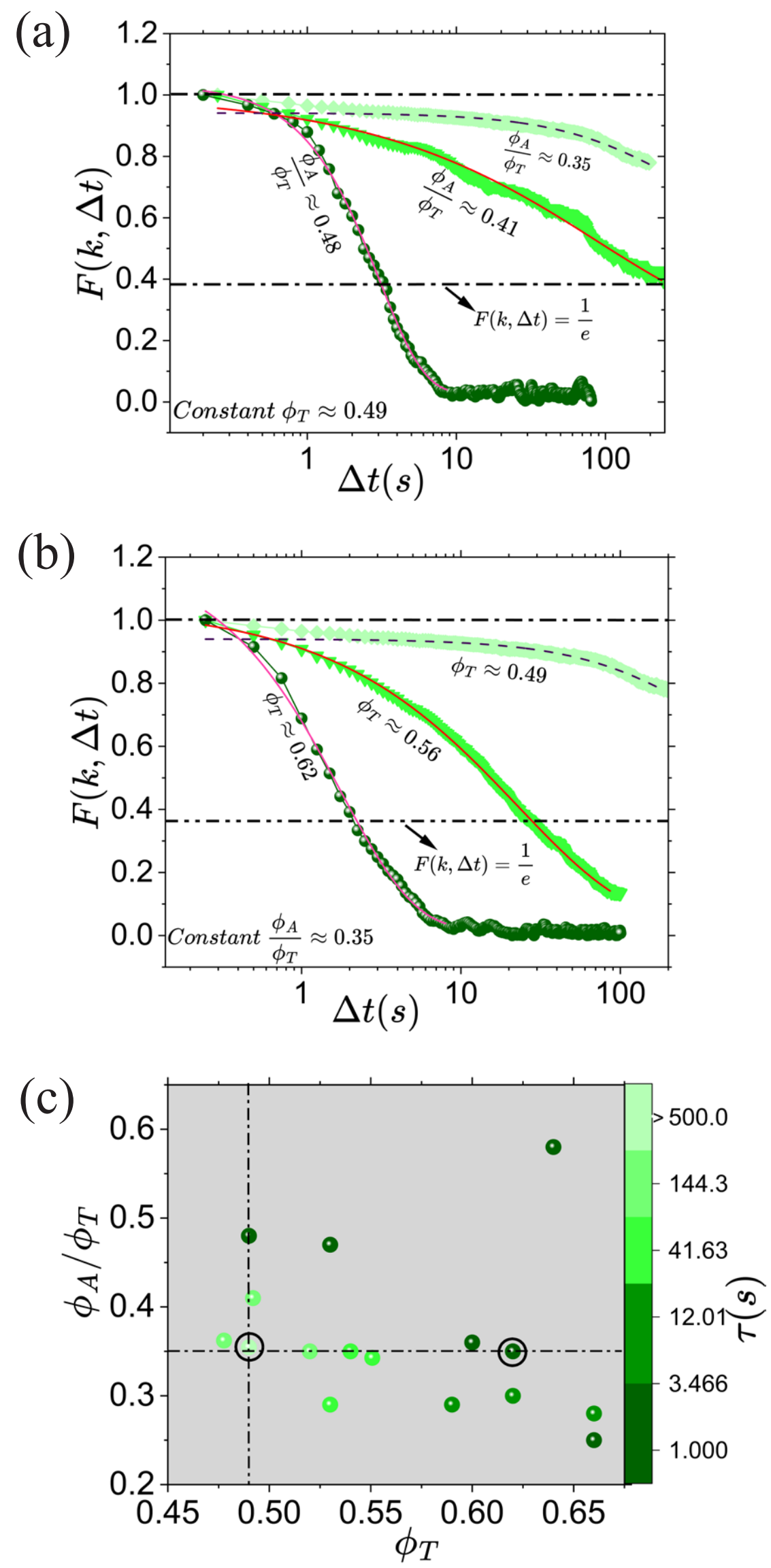

**Figure 4:** Network dynamics. $F(k, \Delta t)$ vs $\Delta t$ for varying $\frac{\emptyset_A}{\emptyset_T}$ at constant $\emptyset_T \sim 0.49$ **(a)** and for varying $\emptyset_T$ at constant $\frac{\emptyset_A}{\emptyset_T} \sim 0.35$ (**b**). Here k ~ 1/r. In a and b, the solid fits are stretched exponential (KWW), i.e.,~ $exp\left[-\left(\frac{t}{\tau}\right)^{\beta}\right]$ and the dashed fit is plateauing decay ~ $exp\left[-\left(\frac{t}{\tau}\right)\right]$ **(c)** Phase map with relaxation time color-coded by τ in the phase space of networks. Dashed lines represent the states shown in Figs. a-b.

The non-identical and hierarchical interparticle interactions lead to distinct local compositions of active and passive particles. To resolve this, we characterize the local environment using the nearest neighbor configuration, $C_f = (N_{nn}, N_{nn}^A)$, which specifies both $N_{nn}$ and the number of active neighbors per passive, $N_{nn}^A$.

$P(C_f)$ reveals a limited set of favored configurations in both slow and fast networks (**Fig 5b and SI 5)**. In these configurations, active particles are separated by at least one passive particle, while passive particles can form nearest neighbors. This is consistent with the radial pair correlation functions (**Fig 5c**):$g_{aa}(r)$ for active particles show no peak at $r/\sigma \sim 1$ and instead peaks at $r/\sigma \sim 2$, indicating separation by an intervening passive particle. In contrast, $g_{pp}(r)$ shows the first peak at $r/\sigma \sim 1$, showing passive–passive nearest-neighbor contacts. This reflects strong active–active repulsion and a weaker passive–passive repulsion (**Fig 1c**).

The peak of $P(C_f)$ shifts from (3,2) at low $\emptyset_T$ (slow network) to (4,2) at high $\emptyset_T$ (fast network), with additional weight at higher coordination, like (6,2) and (6,3). The increase in coordination number reflects a more compact network structure. Consequently, the first peak of $g_{aa}(r)$ shifts to $\sim 1.7\sigma$ from $\sim 1.9\sigma$ at higher $\emptyset_T$, indicating that active particles are brought into closer proximity. As the coordination increases, the structure incorporates more same-species neighbors, enhancing intraspecies repulsion. Combined with the long-range nature of dipolar interactions, this leads to structural frustration, as particles are forced into configurations that are more closely packed than favored by repulsive interactions. This frustration generates local instabilities within the network.

The instabilities manifest as enhanced orientational dynamics of active particles at higher $\emptyset_T$, as captured by the ensemble-averaged angular autocorrelation function $A(\Delta t)$[37], which decays more rapidly, indicating faster decorrelation. Changes in orientation promote cage breaking, leading to propulsion and local rearrangements, facilitating structural reorganization, thereby accelerating network dynamics (**SI 7, SI Mov. 5**). In contrast, at lower $\emptyset_T$, slower orientational decorrelation leads to stable and slowly relaxing networks (**Fig 5d**).

This mechanism is further supported by comparing $P(C_f)$ at a fixed $\emptyset_T \approx 0.49$ for different $\frac{\emptyset_A}{\emptyset_T}$, where faster dynamics correlate with a higher local coordination number despite an identical total number density (**SI 6**). These results establish the key role of microstructure in determining the emergent dynamics of the network.

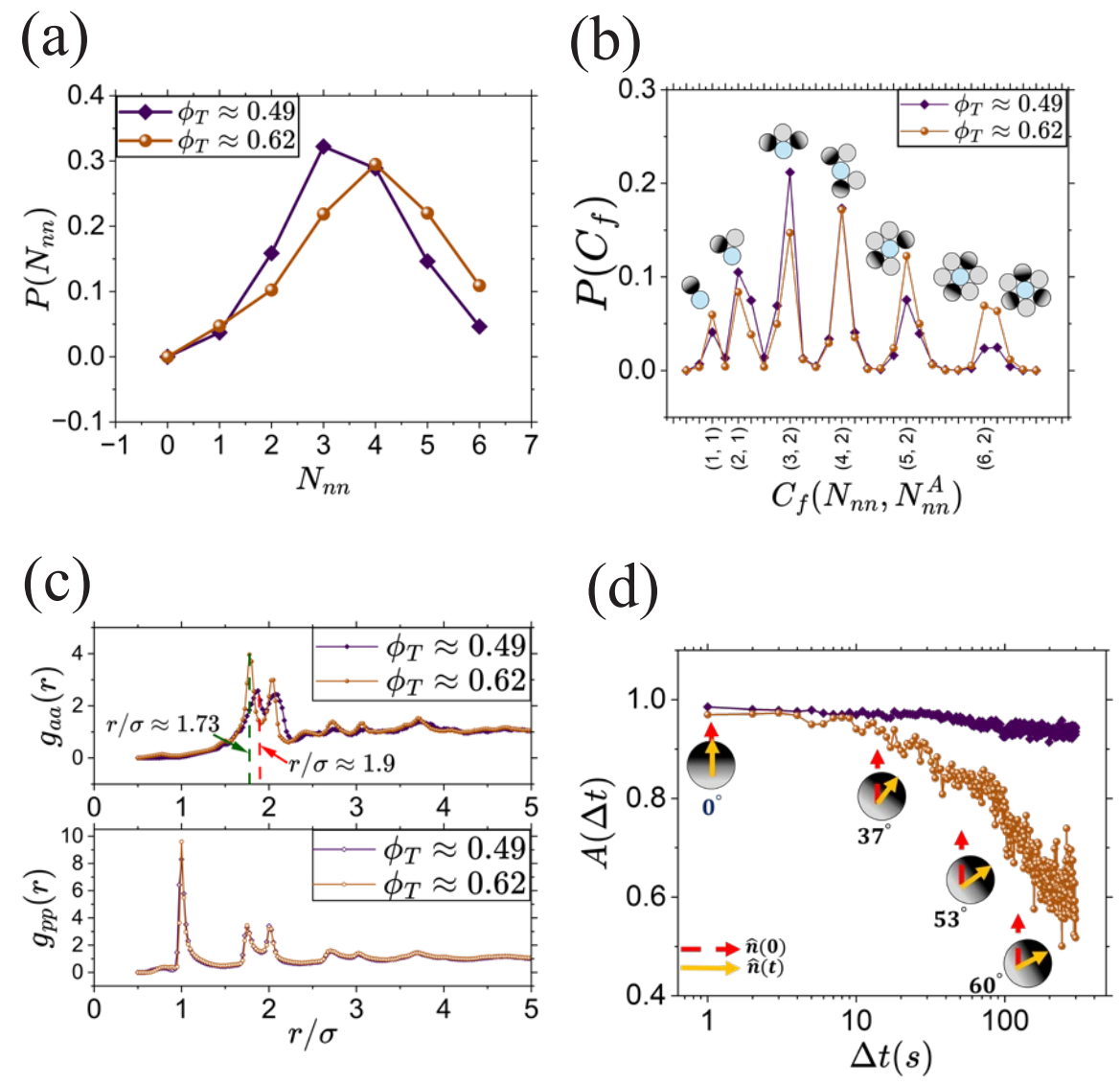

**Figure 5:** Microstructure of the network states at $\frac{\emptyset_A}{\emptyset_T} \sim 0.35$. The states circled in black in Fig 4 **(a)** $P(N_{nn})$ for all the particles in the network. **(b)** $P(C_f)$ with passive particles as the reference particle (alternate representations in SI. 5,6). Schematics of representative configurations are shown for each peak. The reference particles are in blue, passive, and active neighbors shown as grey and Janus spheres, respectively. **(c)** Radial pair correlation function $g_{aa}(r)$ for active (top; 1st peak marked by dashed lines) and for passive ($g_{pp}(r)$) particles, (bottom). **(d)** Angular orientation correlation, $A(\Delta t)$ vs $\Delta t$ for the active particles in the network. Orientation at initial time is $\hat{n}(0)$ shown as red dashed arrow, and $\hat{n}(t)$ for fast network is shown in solid yellow arrows.

In summary, we show that the dynamics and stability of active–passive colloidal networks emerge from the intricate interplay between interparticle interactions and activity-driven rearrangements. Increasing the total area fraction compacts the network, bringing active particles into closer proximity, which amplifies long-range dipolar repulsions and promotes continuous orientational reconfigurations, thereby accelerating structural relaxation. Our microstructural and single-particle analyses reveal that the observed non-trivial dynamics arise from the non-identical dipolar interactions between the active and passive species in addition to activity. More broadly, these results demonstrate that tuning activity and directional interactions provide a route to control network structure and dynamics, offering design principles for reconfigurable active gels.

**Acknowledgements:** The authors thank Bipul Biswas for assisting with the fabrication and optimization of the particles and A. D. Dinsmore for feedback on the work. H.N.K. thanks the University of Massachusetts Amherst for the start-up funding.

------------------------------------------------------------

## End Matter

*Clustering Algorithm:* Clusters were identified using a distance-based connectivity criterion. Particle positions (x, y) were first linked across consecutive frames to obtain single-particle trajectories. Within each frame, a pairwise center-to-center distance was computed between all particles. Two particles were assigned to the same cluster if their separation was smaller than a fixed cutoff of $1.2\sigma$, where $\sigma$ is the particle diameter. The cutoff distance was chosen to be slightly larger than the particle diameter to account for positional uncertainty and thermal fluctuations and does not affect the qualitative trends reported here. Clusters were then defined as connected components of this adjacency criterion, such that all particles connected formed a single cluster. The cluster size ($S_{Clus}$) is defined as the total number of particles in a connected component in each frame [38].

*Dynamical Characterization:* we compute the mean-squared displacement (MSD)(**EM Fig 1a**); $\langle\overline{\delta r^2}\rangle = \langle(r(t) - r(0))^2\rangle \propto t^{\alpha}$ in its limiting form and find that all the observed states exhibit super-diffusive behavior with $\alpha > 1$. For a fixed $\emptyset_T$, $\alpha$ increases with $\frac{\emptyset_A}{\emptyset_T}$ and approaches 2, indicating ballistic motion for the re-entrant transitions. For a fixed $\frac{\emptyset_A}{\emptyset_T}$, $\alpha$ initially decreases with increasing $\emptyset_T$, from fluid to cluster state consistent with crowding. However, comparing MSDs for different network states, we find that for the same $\frac{\emptyset_A}{\emptyset_T}$, $\alpha$ increases with $\emptyset_T$ within the percolated network regime**(EM Fig. 1b)**. The long-time effective diffusion coefficient ($D_{eff}$) extracted from the MSD using the following functional form: $\langle\overline{\delta r^2}\rangle \sim 4D_{eff}t$. Contrary to expectations, dynamics become faster as the networks become denser, as indicated by the increase in $\alpha$ and $D_{eff}$**(EM Fig. 1).** Nevertheless, the Brownian diffusion coefficient ($D_0$) of these states in the absence of the field decreases with $\emptyset_T$ consistent with the slowing down of dynamics with increasing particle number density (**SI 4**).

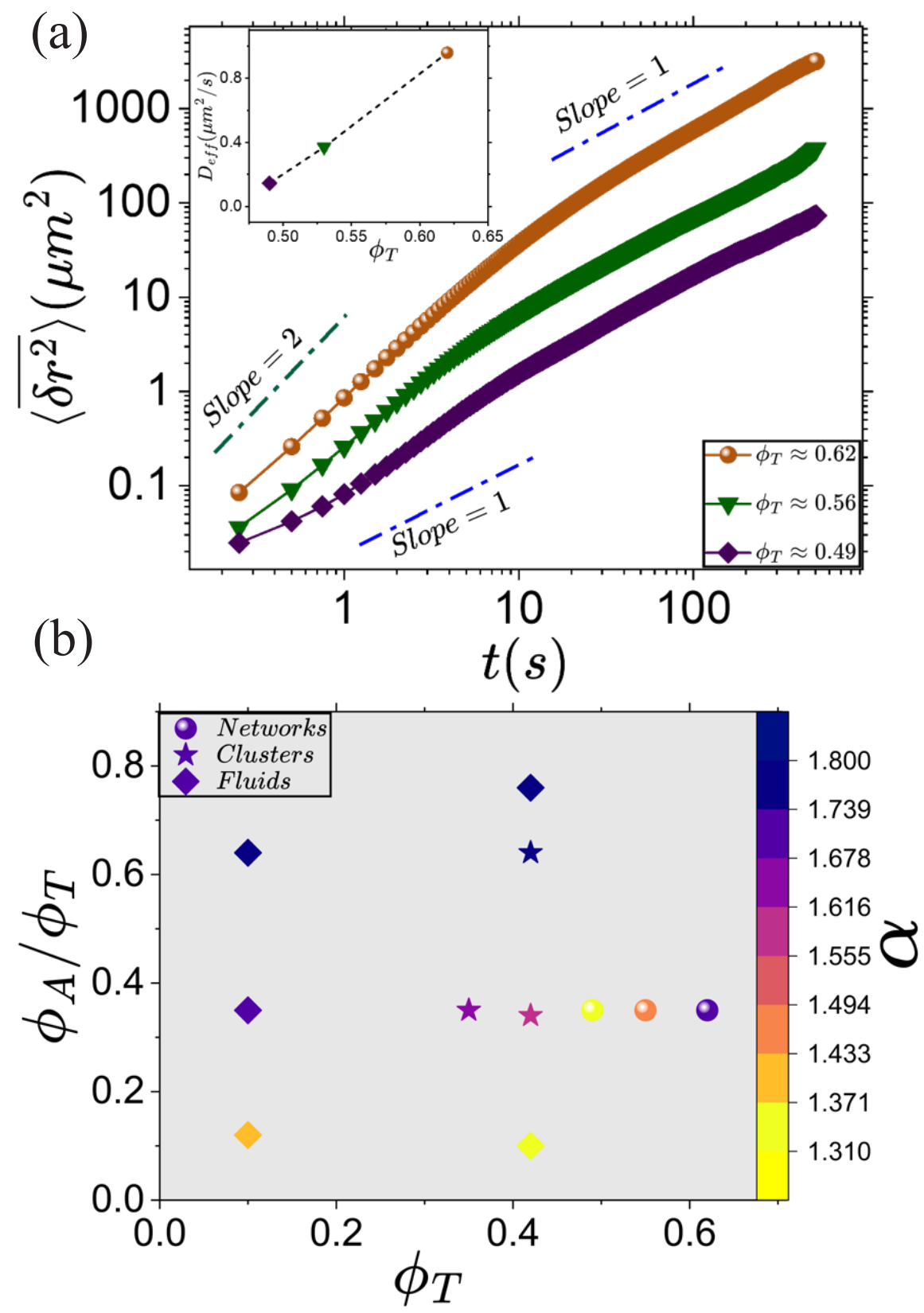

**EM Fig 1: (a)** $\langle\overline{\delta r^2}\rangle$ vs time for particles in percolated network states (high $\emptyset_T$ dense state). **Inset:** long-time diffusion coefficient, $D_{eff}$ vs $\emptyset_T$ for each curve. **(b)** $\frac{\emptyset_A}{\emptyset_T}$ vs $\emptyset_T$ showing the variation of $\alpha$(color-coded) and superdiffusive nature across different states, which was extracted from the MSD ($\langle\overline{\delta r^2}\rangle \sim t^{\alpha}$) of the particles in those states.

We estimate the relaxation time of the networks (**Fig.4**) using the ensemble-averaged self-intermediate scattering function F (k, Δt), given by:

$$F(k, \Delta t) = \langle \frac{1}{N} \sum_{j=1}^{N} e^{ik.[r_j(\Delta t)]} \rangle$$

Where N is the total number of particles in the network, $k \sim \frac{1}{r}$, wavevector corresponding to the first peak of the g(r) and $r_j(\Delta t) = r_j(t) - r_j(0)$ is the difference in particle positions at time t=t and t=0. Relaxation time ($\tau$) of the network is the characteristic time extracted from fitting $F(k, \Delta t)$ with KWW and plateauing decay, and around similar timescales as the raw $F(k, \Delta t)$ decays to $\frac{1}{e}$, which is marked with a dashed line in **Fig 4a-b**.

*Calculation of Orientation of active colloids:* Enhanced orientational fluctuations of active particles in the network are calculated in the ensemble-averaged angular autocorrelation function, averaged over all the frames in steady state, by the following equation:

$$A(\Delta t) = < \hat{n}(\Delta t).\hat{n}(0) >$$

where $\hat{n}(\Delta t)$ is the orientation of an active particle at lag time $\Delta t$ and $\hat{n}(0)$ is the orientation at the initial time during steady-state. To obtain the orientation of active particles in the structure, we conducted 100x microscopy experiments at the same area fractions of slow and fast networks so that the metal and dielectric faces can be clearly analyzed in custom-written MATLAB codes to extract the orientation.

*Role of Hydrodynamics:* To characterize the influence of hydrodynamic flow in dynamics, for the networks in their steady state, we compute the normalized velocity spatial correlation function given by [39]:

$$C_s(R) = \frac{\langle v(r).v(r+R)\rangle_r}{\langle v^2(r)\rangle_r}$$

Where $v$ is the velocity of the particles and $\langle . \rangle$ is the spatial average of all the positions of r. We plot $C_s(R)$ $vs$ $\frac{R}{\sigma}$ , where $\sigma$ is the particle diameter. We obtain the velocity correlation length ($\xi$) for the networks when $C_s(R)$ falls to $\frac{1}{e}$ at different $\emptyset_T$ while keeping the $\frac{\emptyset_A}{\emptyset_T}$ constant at $\approx 0.35$.

**EM Fig 2** shows that although the dynamics of the networks increase gradually with $\emptyset_T$, the correlation length remains constant with increasing $\emptyset_T$ from 0.49 to 0.55, and $\xi \approx 0.8$ particle diameters.

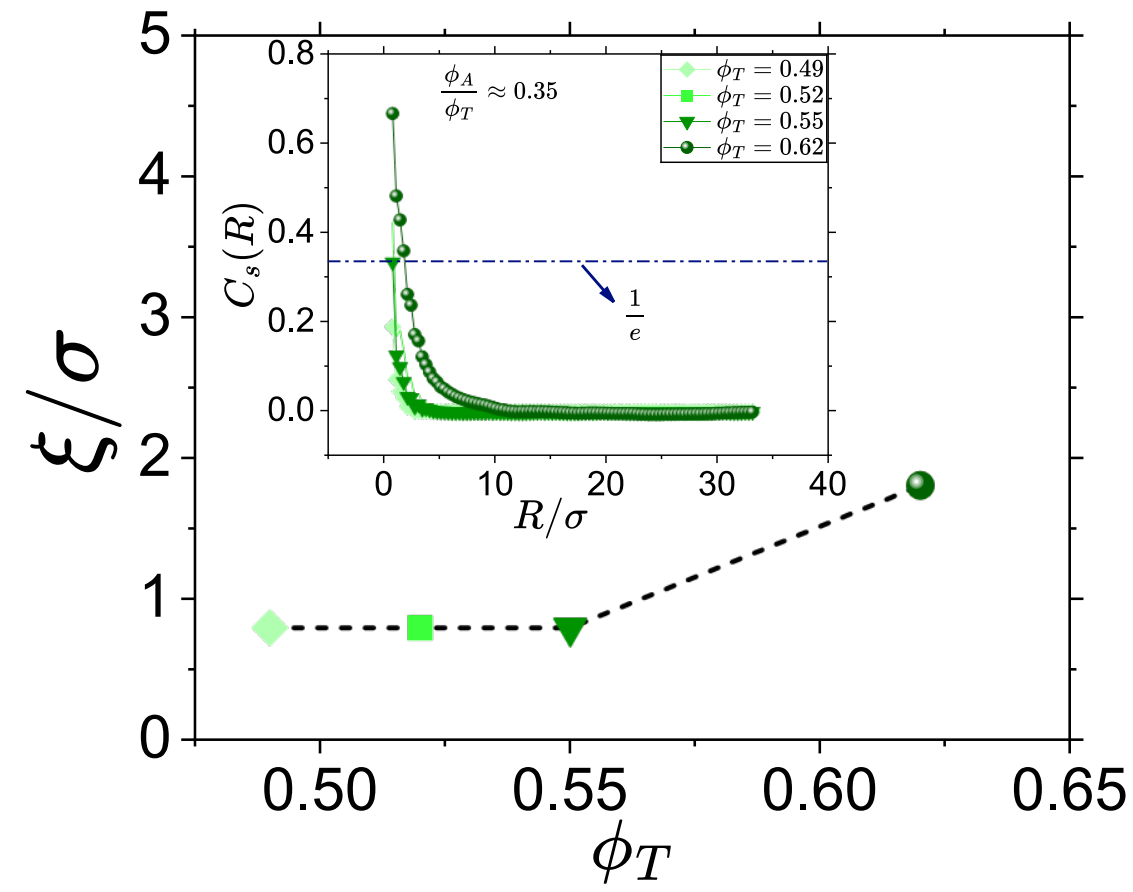

**EM Fig 2:** Correlation length ($\xi$) Vs $\phi_T$ for the networks at a constant $\frac{\phi_A}{\phi_T} \approx 0.35$. $\xi$ is normalized by particle diameter ($\sigma$) **Inset:** $C_s(R)$ Vs R normalized by $\sigma$. Symbols transitioning from lighter to dark shades of green represent the networks from slowest to fastest.

For the densest and fastest sample in our state diagram ( $\emptyset_T \approx 0.62$), $\xi$ slightly increases to $\approx 1.8$ particle diameters, which is quite small. This indicates that any hydrodynamic effects present are short-ranged and lie well within the interaction range extracted from pair measurements (Fig. 1, main manuscript). Moreover, the hydrodynamic length scale estimated here is much smaller than that associated with the electrohydrodynamic flows, which are typically observed at frequencies below 100 kHz [40,41], well below the 1 MHz driving frequency used in our experiments. Thus, although hydrodynamics may contribute, the network properties observed here likely arise primarily from strong interparticle interactions [42].